\documentstyle[12pt]{article}  
 \title{ On the spatial behaviour in dynamics of  porous elastic mixtures
\thanks{Work performed in the context of the COFIN MURST, Italy,
"Mathematical Models for Materials Science"and under auspicies of G. N. F. M. of
the Italian Research Council (C. N. R. ). }} 
 \author{Michele Ciarletta, \quad Gerardo Iovane, \quad Francesca Passarella
 \thanks{Tel.: +39-89-964252, fax:  +39-89-964191
E-mail address: passarella@diima. unisa. it}}  \date{ 
{\small Department of Information Engineering and Applied Mathematics (DIIMA),}
\\
{\small University of Salerno, 84084 Fisciano (Sa) Italy}
}

\def\c#1{\setbox0=\hbox{#1}\ifdim\ht0=1ex \accent'30 #1%
\else{\ooalign{\hidewidth\char'30\hidewidth\crcr\unhbox0}}\fi}

 \def\bull{\vrule height .9ex width .8ex depth -.1ex}

\def \K{ \stackrel{ }{\chi} }

\begin {document}
 
\maketitle

\begin{quote}
{\bf Abstract}- {\small
In this paper we study the spatial and temporal behaviour of the dynamic
processes in  porous elastic mixtures. For the spatial behaviour we use the time-weighted
surface power  function method in order to obtain a more precisely determination of 
the domain of influence and we establish spatial decay estimates of Saint--Venant type with
time-independent  decay rate for the inside of the domain of influence. For the asymptotic temporal behaviour we use the Ces\'{a}ro means associated with the kinetic and strain energies and establish the asymptotic equipartition of the total energy.
 A uniqueness theorem is proved for finite and infinite bodies and 
 we note that it is free of any kind of a priori assumptions of the solutions at infinity. }
\end{quote}

\section{Introduction}

Various theories have been proposed in literature for describing the behaviour of the chemically reacting media (see, for example, Truesdell and Toupin \cite{[M1]}, Kelly \cite{[M2]}, Eringen and Ingram 
\cite{[M3],[M4]}, Green and Naghdi \cite{[M5],[M6]}, M\" {u}ller \cite{[M7]}, Dunwoody and M\"{u}ller \cite{[M8]}, Bedford and Drumheller \cite{[M9]}, etc).

Recently Ie\c {s}an \cite{[M10]} has developed a theory for binary mixtures of granular materials in Lagrangian description in which the independent constitutive variables are the displacement gradients, displacement fields, volume fractions and volume fraction gradients. The theory takes into account the results established previously by Nunziato and Cowin \cite{[M11]}, Goodman and Cowin \cite{[M12]} and Drumheller \cite{[M13]}. The intended applications for such a theory are to granular composites, solid explosives and geological materials.

In \cite{[M10]} a linear theory is also presented and some uniqueness results for bounded bodies are established for the linear dynamic theory with no definiteness assumptions on the elasticities and without any restriction on the initial stresses.

The present paper studies the spatial and temporal behaviour of the solutions to the boundary--initial value problems in the linear dynamic theory of porous elastic mixtures as developed in \cite{[M10]}.

For the spatial behaviour of the dynamic processes in porous elastic mixtures we use the time--weighted surface power method developed in \cite{[M14]}. Thus, we introduce a time--weighted surface measure associated with the dynamic process in question and then we establish a first--order partial differential inequality whose integration gives a good information upon the spatial behaviour. Then we obtain a more precisely version of the domain of influence in the sense that for each fixed $t\in[0,T]$ the whole activity is vanishing at distances to the support of the given data on $[0,T]$ greater than $ct$, where $c$ is a constant characteristic to the elastic mixture. A spatial decay estimate of Saint--Venant's type is established for describing the spatial behaviour of the dynamic process inside of the domain of influence.

As regards the temporal behaviour of the dynamic processes in porous elastic mixtures, we introduce the Ces\'{a}ro means of various energies and then establish the relations describing the asymptotic equipartition of energy. In this aim we use some Lagrange identities and the method developed by Day \cite{[M15]} and Levine \cite{[M16]}.

The plan of our paper is the following one. In the Section 2 we present the basic equations of the linear dynamic theory of porous mixtures developed in \cite{[M10]}. Some constitutive assumptions and other useful results are also presented. The auxiliary identities are established in the Section 3, while in the Section 4 a time--weighted surface measure is defined and its properties are studied. Moreover, a first--order partial differential inequality is established for this measure. The main result concerning the spatial behaviour is presented in Section 5 and some uniqueness results are obtained as a direct consequence. In the Section 6 we introduce the Ces\'{a}ro means of various energies and establish the asymptotic equipartition of the total energy.

\section {Basic equations}

Throughout this article, the motions of continuum are
studied respect to a fixed orthonormal  frame in $ {\rm
I}\!{\rm R}^3$. Then, we deal with functions of  position
and time. Moreover, it is useful stress that in the
following text the tensor components  of order $p\geq
1$ will appear with Latin subscripts, ranging over the
integers $\{1, 2, 3\}$, and summation over repeated
subscripts will be implied. 
Greek indices are understood to range over $\{1, \dots,
9\}$  if they are lower case letters, or over $\{1, 2 \}$ 
if they are upper case letters; summation convention is
not used for these indices. Occasionally, we shall use
bold-face character and typical notations for vectors and
operations upon them. Superposed dots or subscripts
preceded by a comma will mean partial derivative with
respect to the time or the corresponding coordinates.

Let $ B $ be  a bounded or unbounded
 regular region  in the physical $3$--dimensional space, whose boundary  $\partial B$
is a piecewise smooth surface.
A chemically inert binary 
mixture of two interacting porous  elastic solids,
$c_1$ and $c_2$, in a given 
 re\-ference configuration, is into B.

The positions of particles of 
$c_1$ and $c_2$ at time t are ${\bf x}$ and ${\bf y}$ respectively, i. e. 
$$
{\bf x}={\bf x}({\bf X}, t), \qquad {\bf y}=
{\bf y}({\bf Y},t)
\qquad\quad {\bf X},{\bf Y}\in B,\; t\in I,
$$ 
in which ${\bf X}$ and ${\bf Y}$ are reference positions of
 these particles, $ I=[0, \infty)$. 
By following Bedford and Stern \cite{[M17]}, we assume that ${\bf X}=
{\bf Y}$. 

Let the top label $\alpha$ refer the various fields to the
constituent $c_{\alpha}$. Taking into account the linear theory,
the behaviour of a binary mixture of elastic solids is governed by the 
local balance equations (see  Iesan \cite{[M10]})
\medskip

\begin{equation}
\begin{array}{l}
S_{ji,j}^{(\alpha)} +( -1 )^{\alpha} p_i + \rho^{(\alpha)} f_i^{(\alpha)}= \rho^{(\alpha)}
\ddot {u}_i^{(\alpha)},
\\[5 mm]
h_{i,i}^{(\alpha)} +g^{(\alpha)} +\varrho^{(\alpha)}
\ell^{(\alpha)}= \rho^{(\alpha)}  \K^{(\alpha)}\ddot
{\varphi}^{(\alpha)},
\qquad \qquad \hbox{ on }B\times (0, +\infty)
\end{array}
\label{2. 3}
\end{equation}
\smallskip

In these equations, 
$
{\bf S}^{(\alpha)}, {\bf f}^{(\alpha)} $ are the stress
tensor and body force associated to $c_\alpha$;
${\bf p}$ is the vector field for characterising the
mechanical  interaction between the constituents $ c_1
$ and $ c_2$;
$ {\bf h}^{(\alpha)},  g^{(\alpha)}, 
\ell^{(\alpha)}$are the equilibrated stress vector,
intrinsic  and extrinsic equilibrated body force 
associated to $c_\alpha$,  respectively.

Moreover, ${\bf u}^{(\alpha)} $ is the displacement vector fields
associated to $c_\alpha$; $\varphi^{(\alpha)} $ is the
changes in volume fraction
 starting from the reference configuration to $c_\alpha$.

 Finally,  $\rho^{(\alpha)},  \K^{(\alpha)}$ are the
 bulk mass density
and equilibrated inertia of the material $c_\alpha$ 
in the reference state.

According to classical interpretation of system (\ref{2. 3}), we assume that 
\newline
i. \,\,\,
$ u_i^{(\alpha)}, \varphi^{(\alpha)} \; 
\in C^{2, 2}(\bar{B}\times I)$;
\newline
ii. \,\,
$ 
 S_{ji}^{(\alpha)}, h_i^{(\alpha)} \in C^{1, 0}(\bar{B}\times I), 
\; p_i \in C^{0,0}(\bar{B}\times I)$; \newline
iii. \,
$ f_i^{(\alpha)}, g^{(\alpha)}, \ell^{(\alpha)}
 \in C^{0, 0}(\bar{B}\times I), \; 
\rho^{(\alpha)},  \K^{(\alpha)} 
 \in C^{0}(\bar{B})$, 
\newline
where, $\bar B$ is the closure of $B$.

Then, we introduce the 29-dimensional vector field 
 $${\bf E}({\bf U})\equiv \{e_{ij}({\bf U}),
 g_{ij}({\bf U}), 
\varphi^{(1)},\varphi^{(2)},
d_{i} ({\bf U}), \varphi^{(1)}_{,i}, \varphi^{(2)}_{,i}\},$$
with 
$${\bf U}\equiv \{{\bf
u}^{(1)},{\bf u}^{(2)}, \varphi^{(1)},\varphi^{(2)}\}
$$
 and
 \begin{equation}
\begin{array}{lr}
e_{ij}({\bf U})=\frac 12(u_{i,j}^{(1)}+u_{j,i}^{(1)}), \quad \quad
 g_{ij}({\bf U}) =u_{j,i}^{(1)}+u_{i,j}^{(2)},& \\[5 mm]
 d_i({\bf U})=u_i^{(1)} -u_i^{(2)}, 
&\quad \hbox{ on }\bar B\times I. 
\end{array}
\label{geom}\end{equation}
Now, we define the magnitude of ${\bf E}({\bf U})$  by
$$
\begin{array}{l}
\displaystyle
|{\bf E}({\bf U})|\equiv \Bigl\{
\sum_{\alpha=1}^{2} \Bigl[e_{ij}({\bf U}) e_{ij} ({\bf
U})+g_{ij}({\bf U}) g_{ij} ({\bf U})
+\varphi^{(\alpha)} ({\bf U}) \varphi^{( \alpha)} ({\bf
U})+
\\[5 mm]
\displaystyle
\qquad \qquad
+d_i^2 ({\bf U}) 
+
\varphi_{,i}^{(\alpha)} ({\bf U}) \varphi_{,i}^{( \alpha)} ({\bf U})
\Bigr] \Bigr\}^{1/2}. 
\end{array}
$$

Our attention is focused on homogeneous, centrosymmetric mixture, 
by supposing the initial continuum is free from stresses. 

In the context of our theory, the internal energy density associated to ${\bf U}$ 
is given by 
\begin{equation}
\begin{array}{rl}
W({\bf U})=&\displaystyle
 \frac12 \Bigl[
 A_{ijrs} e_{ij}({\bf U}) e_{rs}({\bf U})+
 C_{ijrs} g_{ij}({\bf U}) g_{rs}({\bf U})+
\zeta \varphi^{(1)}\varphi^{(1)} + 
\\[5 mm]
&+  
\mu  \varphi^{(2)}\varphi^{(2)} + 
\alpha_{ij} \varphi^{(1)}_{,i}\varphi^{(1)}_{,j} +
\gamma_{ij} \varphi^{(2)}_{,i}\varphi^{(2)}_{,j} 
+a_{ij}d_i ({\bf U})d_j({\bf U})\Bigr]+
\\[5 mm]
&
+ B_{ijrs} e_{ij}({\bf U}) g_{rs}({\bf U})
+ D_{ij} e_{ij}({\bf U})\varphi^{(1)}+ 
E_{ij} e_{ij}({\bf U})\varphi^{(2)}+
\\[5 mm]
&
+ M_{ij} g_{ij}({\bf U})\varphi^{(1)}+ 
N_{ij} g_{ij}({\bf U})\varphi^{(2)}+
\beta_{ij} \varphi^{(1)}_{,i}\varphi^{(2)}_{,j}+
\\[5 mm]
&
+
b_{ij} d_i({\bf U})\varphi^{(1)}_{,j}+
c_{ij} d_i({\bf U})\varphi^{(2)}_{,j}+
\tau \varphi^{(1)}\varphi^{(2)}. 
\end{array}
\label{2. 5}
\end{equation}
The material coefficients, appearing in previous 
equations (\ref{2. 5}), are constants and they obey
the following symmetry relations:
 \begin{equation}
\begin{array}{l}
A_{ijrs}= A_{jirs}= A_{rsij},\quad
B_{ijrs} =B_{jirs},\quad C_{ijrs}= C_{rsij}, 
\quad a_{ij}=a_{ji}, \\[5 mm]
 \alpha_{ij}=\alpha_{ji}, \quad 
\gamma_{ij}=\gamma_{ji}, \quad 
D_{ij}=D_{ji}, \quad 
E_{ij}=E_{ji}. 
\end{array}
 \label{2. 4}
 \end{equation}
The constitutive equations are
 \begin{equation}
\begin{array}{l}	
S_{ji}^{(1)} ({\bf U}) =  
(A_{jirs}+ B_{rsji}) e_{rs}({\bf U}) +
(B_{ijrs}+ C_{jirs}) g_{rs}({\bf U}) +\\[5 mm]
\qquad +(D_{ij}+ M_{ij})\varphi^{(1)} +
(E_{ij}+ N_{ij})\varphi^{(2)},
\\[5 mm]
S_{ji}^{(2)}({\bf U})  = 
B_{rsij} e_{rs}({\bf U}) +
 C_{ijrs} g_{rs}({\bf U}) 
+M_{ij}\varphi^{(1)} +N_{ij}\varphi^{(2)},
\\[5mm]
g^{(1)}  ({\bf U}) =-D_{rs}e_{rs}
 -M_{rs}g_{rs} -\zeta\varphi^{(1)} -
\tau\varphi^{(2)},
\\[5mm]
g^{(2)}  ({\bf U}) = -E_{rs}e_{rs}
 -N_{rs}g_{rs} -\tau\varphi^{(1)} -
\mu\varphi^{(2)},
\\[5 mm] 
p_i ({\bf U}) =a_{ij}d_j({\bf U}) + 
b_{ij}\varphi^{(1)}_{,j}+c_{ij}\varphi^{(2)}_{,j},
\\[5mm]
h_i^{(1)}  ({\bf U})
=\alpha_{ij}\varphi^{(1)}_{,j}+\beta_{ij}\varphi^{(2)}_{,j}+b_{ji}d_j({\bf U}), 
\\[5mm]
 h_i^{(2)} ({\bf U})  =\beta_{ji}\varphi^{(1)}_{,j}+
\gamma_{ij}\varphi^{(2)}_{,j}+c_{ji}d_j({\bf U}).
 \end{array}
 \label{2. 2} 
\end{equation}

Let ${\cal A}_1=\| \tilde a_{KL}\| \;\, 
(K,L= 1, \dots, 20)  $  
\begin{equation}
 \begin{array}{llll}
\tilde a_{\Gamma\, \Delta}=A_{\Gamma\,\Delta}, 
&\; \tilde a_{\Gamma\, (9+\Delta)}=B_{\Gamma\,\Delta},
&\; \tilde a_{\Gamma\, 19}=D_\Gamma,
&\;  \tilde a_{\Gamma\, 20}=E_\Gamma,
\\[5 mm]
\tilde a_{(9+\Gamma)\, \Delta}=B_{\Delta\,\Gamma}, 
&\; \tilde a_{(9+\Gamma)\,
(9+\Delta)}=C_{\Gamma\,\Delta}, 
&\; \tilde a_{(9+\Gamma)\,
19}=M_\Gamma,
 &\;  \tilde a_{(9+\Gamma)\, 20}=N_\Gamma,
\\[5 mm]
\tilde a_{19\, \Delta}=D_{\Delta}, 
&\; \tilde a_{19\, (9+\Delta)}=M_{\Delta},
&\; \tilde a_{19\, 19}=\zeta,
&\;  \tilde a_{19\, 20}=\tau,
\\[5 mm]
\tilde a_{20\, \Delta}=E_{\Delta}, 
&\; \tilde a_{20\, (9+\Delta)}=N_{\Delta},
&\; \tilde a_{20\, 19}=\tau,
&\;  \tilde a_{20\, 20}=\mu,
\end{array}
\label{matrix1}
\end{equation}
where we have  called the nine index combinations $(i\,j)$ or $(r\,s)$
 by  capital greak letters $(i.e. \Gamma\,,\,\Delta\,,$ and so on$).$ 
Now, let 
${\cal O}$ be the empty  matrix $20 \times 9$ and 
 ${\cal A}_2=\| \tilde b_{KL}\| \;\, (K,L= 1,
\dots, 9)  $  be
 \begin{equation}
 \begin{array}{lll}
\tilde b_{ij}=a_{ij}, 
&\quad \tilde b_{i\, (3+j)}=b_{ij},
&\quad \tilde b_{i\, (6+j)}=c_{ij},
\\[5 mm]
\tilde b_{(3+i)j}=b_{ji}, 
&\quad \tilde b_{(3+i)\, (3+j)}=\alpha_{ij},
&\quad \tilde b_{(3+i)\, (6+j)}=\beta_{ij},
\\[5 mm]
\tilde b_{(6+i)j}=c_{ji}, 
&\quad \tilde b_{(6+i)\, (3+j)}=\beta_{ji},
&\quad \tilde b_{(6+i)\, (6+j)}=\gamma_{ij}.
\\[5 mm]
\end{array}
\label{matrix2}
\end{equation}

Then, the energy density (\ref{2. 5}) assumes the form 
\begin{equation}
W({\bf U})=\frac{1}{2}\sum_{K,L=1}^{29}\tilde A_{KL}E_K({\bf U}) E_L({\bf U})=
\frac{1}{2}
{\bf E}({\bf U})\cdot {\cal A}{\bf E}({\bf U}), 
\label{stella}
\end{equation}
where the matrix ${\cal A}=\| \tilde A_{KL}\| \;\, (K, L=1,\dots, 29)  $ is defined
by
\begin{equation}
{\cal A}= 
\left[
\begin{array}{ll}
{\cal A}_1  &\quad {\cal O}   \\[3 mm]
{\cal O}^T  &\quad {\cal A}_2 
\end{array}\right],
\label{matrix}
\end{equation}
and  ${\cal O}^T$ is the  transposed matrix of ${\cal
O}$.

In what follows, we assume that $\rho^{(\alpha)}$, $  \K^{(\alpha)}$ 
 are strictly positive and
$W({\bf U})$ is a positive  definite quadratic form; 
thus, there exist the positive constants $\xi_m$ and $\xi_M$ so that
\begin{equation}
\xi_m |{\bf E}({\bf U})|^2\geq   2W({\bf U}) \geq  \xi_M |{\bf E}({\bf U})|^2
\label{ste}
\end{equation}
where $\xi_m$ is the minimum  elastic modulus 
 and $\xi_M$
is the maximum elastic moduli.

Let
$${\bf S}({\bf U})\equiv \{ S_{ji}^{(1)} ({\bf U}), S_{ji}^{(2)} 
 ({\bf U}),g^{(1)} ({\bf U}), g^{(2)} ({\bf U}), p_i ({\bf U}), 
h_{i}^{(1)} ({\bf U}), h_{i}^{(2)}({\bf U})\},$$  
 then the magnitude of $S({\bf U})$ is defined by
$$
|{\bf S}({\bf U})|\equiv \Bigl\{
\sum_{\alpha=1}^{2} \Bigl[S_{ji}^{(\alpha)} ({\bf U}) S_{ji}^{( \alpha)} ({\bf
U})+
h_{i}^{(\alpha)} ({\bf U}) h_{i}^{( \alpha)} ({\bf
U})
+g^{(\alpha)} ({\bf U}) g^{( \alpha)} ({\bf
U})
\Bigr]+  p_i ({\bf U}) p_{i} ({\bf U})\}^{1/2}. 
$$
Taking into account the equations
 (\ref{2. 2}, \ref{matrix1}, \ref{matrix2}, \ref{stella}, \ref{matrix}),
we prove that 
\begin{equation}
	|{\bf S}({\bf U})|^2 =  {\cal A}{\bf E}\cdot {\cal A}{\bf E}=
{\bf E}\cdot {\cal A}^2{\bf E}
\leq \xi_M  {\bf E}\cdot {\cal A}{\bf E}= 2\xi_M  W({\bf U}). 
\label{ok}
\end{equation}

The surface tractions  ${\bf s}^{(\alpha)}({\bf U})$ and
 $h^{(\alpha)}({\bf U})$ are defined by
\begin{equation}
s_i^{(\alpha)}({\bf U}) = S_{ji}^{(\alpha)}({\bf U})n_j, \qquad
h^{(\alpha)}({\bf U}) = h_{j}^{(\alpha)}({\bf U})n_j,
 \label{2. 9}
 \end{equation} 
where ${\bf n}$ is the
outward unit normal  vector to boundary surface. 
The relations (\ref{ok}, \ref{2. 9}) imply that
\begin{equation}
\sum_{\alpha=1}^{2} [s_{i}^{(\alpha)} ({\bf U}) s_{i}^{( \alpha)} ({\bf U})
+h^{(\alpha)} ({\bf U}) h^{( \alpha)} ({\bf U})
]
\leq |{\bf S}({\bf U})|^2\leq 2\xi_M  W({\bf U}). 
\label{okok}
\end{equation}

If we introduce the notations
\begin{equation}
\begin{array}{l}
a_{ijrs}= A_{jirs}+ B_{rsji }+ B_{jisr }+C_{jisr},\\[5 mm]
 b_{ijrs} =B_{jirs}+ C_{jirs},
\qquad \qquad
d_{ijrs}= C_{ijrs},
\\[5 mm]
 \tau_{ij} =D_{ij}+ M_{ij}
\qquad \qquad
 \sigma_{ij} =E_{ij}+ N_{ij},
\end{array}
 \label{*}
 \end{equation}
 the equations (\ref{2. 5}) become
\begin{equation}
\begin{array}{rl}
W({\bf U})=&\displaystyle
 \frac12 \Bigl[
 a_{ijrs} u_{i,j}^{(1)} u_{r,s}^{(1)}+
 d_{ijrs}u_{i,j}^{(2)} u_{r,s}^{(2)}+
\zeta \varphi^{(1)}\varphi^{(1)} + 
\mu  \varphi^{(2)}\varphi^{(2)} + 
\\[5 mm]
&+  
\alpha_{ij} \varphi^{(1)}_{,i}\varphi^{(1)}_{,j} +
\gamma_{ij} \varphi^{(2)}_{,i}\varphi^{(2)}_{,j} 
+a_{ij}d_i ({\bf U})d_j({\bf U})\Bigr]+b_{ijrs}  u_{i,j}^{(1)} u_{r,s}^{(2)}
+ 
\\[5 mm]
&
+\tau_{ij}  u_{i,j}^{(1)} \varphi^{(1)}+ 
 \sigma_{ij}  u_{i,j}^{(1)} \varphi^{(2)}
+ M_{ij} u_{i,j}^{(2)}\varphi^{(1)}
+ 
N_{ij} u_{i,j}^{(2)}\varphi^{(2)}+  \\[5 mm]
&
+
\beta_{ij} \varphi^{(1)}_{,i}\varphi^{(2)}_{,j}
+
b_{ij} d_i({\bf U})\varphi^{(1)}_{,j}+
c_{ij} d_i({\bf U})\varphi^{(2)}_{,j}+
\tau \varphi^{(1)}\varphi^{(2)}. 
\end{array}
\label{2. 5bis}
\end{equation}
Using the symmetry relation (\ref{2. 4}), we get 
\begin{equation}
\begin{array}{l}
a_{ijrs} = a_{rsij}
,\qquad \qquad d_{ijrs} = d_{rsij}
,\qquad \qquad
a_{ij}=a_{ji}.
\\[5 mm]
\alpha_{ij}=\alpha_{ji}
,\qquad \qquad\gamma_{ij}=\gamma_{ji}
. 
\label{2. 4bis}
\end{array}
\end{equation}

The constitutive equations  (\ref{2. 2}) become
 \begin{equation}
\begin{array}{l}	
S_{ji}^{(1)} ({\bf U}) = a_{ijrs} u_{r,s}^{(1)}+
b_{ijrs} u_{r,s}^{(2)} +\tau_{ij}\varphi^{(1)} +\sigma_{ij}\varphi^{(2)},
\\[5 mm]
S_{ji}^{(2)}({\bf U})  = b_{rsij} u_{r,s}^{(1)}+
d_{ijrs} u_{r,s}^{(2)}+M_{ij}\varphi^{(1)} +N_{ij}\varphi^{(2)},
\\[5mm]
g^{(1)}  ({\bf U}) =-\tau_{rs}u_{r,s}^{(1)} 
 -M_{rs}u_{r,s}^{(2)} -\zeta\varphi^{(1)} -
\tau\varphi^{(2)},
\\[5mm]
g^{(2)}  ({\bf U}) = -\sigma_{rs}u_{r,s}^{(1)} 
 -N_{rs}u_{r,s}^{(2)} -\tau\varphi^{(1)} -
\mu\varphi^{(2)},
\\[5 mm] 
p_i ({\bf U}) =a_{ij}d_j({\bf U}) + 
b_{ij}\varphi^{(1)}_{,j}+c_{ij}\varphi^{(2)}_{,j},
\\[5mm]
h_i^{(1)}  ({\bf U})
=\alpha_{ij}\varphi^{(1)}_{,j}+\beta_{ij}\varphi^{(2)}_{,j}+b_{ji}d_j({\bf U}), 
\\[5mm]
 h_i^{(2)} ({\bf U})  =\beta_{ji}\varphi^{(1)}_{,j}+
\gamma_{ij}\varphi^{(2)}_{,j}+c_{ji}d_j({\bf U}).
 \end{array}
 \label{2. 2bis} 
\end{equation}

It follows from the equations
 (\ref{2. 5bis}), (\ref{2. 4bis}), (\ref{2. 2bis}) that
\begin{equation}
2 W({\bf U})=\displaystyle \sum_{\alpha=1}^2
\Bigl[S_{ji}^{(\alpha)} ({\bf U})u_{i,j}^{(\alpha)}+
p_i({\bf U})d_i ({\bf U}) + h_i^{(\alpha)}({\bf U})\varphi^{(\alpha)}_{,i}  - 
g^{(\alpha)}({\bf U})\varphi^{(\alpha)}\Bigr]. 
\label{nodotW}
\end{equation}
and
\begin{equation}
\dot W({\bf U})=\displaystyle \sum_{\alpha=1}^2
\Bigl[S_{ji}^{(\alpha)} ({\bf U})\dot u_{i,j}^{(\alpha)}+
p_i({\bf U})\dot d_i ({\bf U}) + h_i^{(\alpha)}({\bf U})\dot \varphi^{(\alpha)}_{,i}  - 
g^{(\alpha)}({\bf U})\dot \varphi^{(\alpha)}\Bigr].
\label{dotW}
\end{equation}

We consider the initial-boundary value problem ${\cal P}$
defined by
the equations of motion (\ref{2. 3}), the geometrical
equations (\ref{geom}) and the constitutive equations 
 (\ref{2. 2}) and the following 
initial-boundary conditions 
\begin{equation}
\begin{array}{l}
u_i^{(\alpha)}=a_i^{(\alpha)}
,\quad \quad 
\dot u_i^{(\alpha)}=\dot a_i^{(\alpha)}
,\quad \quad 
\\[5 mm]
\varphi^{(\alpha)}= \varphi_0^{(\alpha)}
,\quad \quad 
\dot \varphi^{(\alpha)}= \dot \varphi_0^{(\alpha)}
\qquad \qquad \hbox{ on }B\times \{0\}.
\end{array}
\label{inizio}
\end{equation}
and
\begin{equation}
\begin{array}{l}
u_i^{(\alpha)}=\tilde u_i^{(\alpha)}
\quad \hbox{ on } \bar \Sigma_1 \times I,
\qquad 
 s_i^{(\alpha)}=\tilde s_i^{(\alpha)}
\quad \hbox{ on }  \Sigma_2 \times I ,
\\[5 mm]
\varphi^{(\alpha)}= \tilde \varphi^{(\alpha)}
\quad \hbox{ on } \bar \Sigma_3 \times I,
\qquad
h^{(\alpha)}= \tilde h^{(\alpha)}
\quad \hbox{ on }  \Sigma_4 \times I,
\end{array}
\label{bordo}
\end{equation}
where $\Sigma_i$ ($i=1,...,4$) are  the subsets of 
$\partial B$ such that
$$
\bar \Sigma_1 \cup\Sigma_2 = \bar \Sigma_3\cup\Sigma_4 = 
\partial B,
\qquad 
\Sigma_1 \cap\Sigma_2 =  \Sigma_3\cap\Sigma_4 = \emptyset.
$$ 

 The terms on right-hand in equations (\ref{inizio})
and (\ref{bordo}) are prescribed continuous functions;
along with  
 $ {\bf f}^{(1)}, {\bf f}^{(2)},$
 $ \ell^{(1)}, \ell^{(2)}$ these  constitute the
external data of the problem ${\cal P}$. 

An array field ${\bf U}= \{{\bf u}^{(1)}, {\bf u}^{(2)}, \varphi^{(1)}, \varphi^{(2)}\}$, meeting all equation
(\ref{2. 3}, \ref{geom}, \ref{2. 2}, \ref{inizio})
and (\ref{bordo}), will be referres to as a
(regular) 
solution of the problem ${\cal P}$.

\section{Auxiliary identities}

In this section we establish some integral  identities that we will use in next sections.

{\bf Lemma 3.1}    Let ${\bf U}$ be a solution of 
initial-boundary-value problem ${\cal P}$. Then, for every regular region $P\subset B$ with regular boundary 
$\partial P$, it follows that 
\begin{equation}
\begin{array}{l}
\displaystyle
\frac 12
\int_P  
e^{-\lambda t} \sum_{\alpha=1}^2
\bigl[ \rho^{(\alpha)}  \dot u^{(\alpha)}_i(t)\dot u^{(\alpha)}_i(t) +
  \rho^{(\alpha)}   \K^{(\alpha)}  \dot \varphi^{(\alpha)} (t)\dot\varphi^{(\alpha)} (t) 
 + 2W({\bf U}(t)) \bigr] dv +
\\[5 mm] \quad\displaystyle
+ \frac{\lambda}{2}\int_0^t\int_P  e^{-\lambda s}\sum_{\alpha=1}^2
 \bigl[ \rho^{(\alpha)}  \dot u^{(\alpha)}_i(s)\dot u^{(\alpha)}_i(s) +   \rho^{(\alpha)}   \K^{(\alpha)}  \dot \varphi^{(\alpha)}(s) \dot \varphi^{(\alpha)}(s) +
\\[5 mm] \quad  \displaystyle
 + 2 W({\bf U}(s))\bigr] dv ds =
\int_0^t \int_P e^{-\lambda s} \sum_{\alpha=1}^2 [ \rho^{(\alpha)} f^{(\alpha)}_i(s) \dot u^{(\alpha)}_i(s) +  \rho^{(\alpha)}  \ell^{(\alpha)} (s) \dot \varphi^{(\alpha)} (s)] dv ds
+
\\[5 mm] \quad
\displaystyle +
\frac12  \int_P \sum_{\alpha=1}^2 
\bigl[  \rho^{(\alpha)}\dot u^{(\alpha)}_i(0) \dot u^{(\alpha)}_i(0) +
   \rho^{(\alpha)}   \K^{(\alpha)}  \dot \varphi^{(\alpha)} (0)^2  + 2W({\bf U}(0)) \bigr] dv +
\\[5 mm] \quad
\displaystyle 
+\int_0^t \int_{\partial P} e^{-\lambda s} \sum_{\alpha=1}^2[  s^{(\alpha)}_i({\bf U}(s)) \dot
 u^{(\alpha)}_i(s) + h^{(\alpha)}({\bf U}(s)) \dot \varphi^{(\alpha)} (s) ] da ds  , 
\end{array}
\label{A16}
\end{equation}
where $\lambda$ is a positive parameter and  $t\in I$.

 {\bf Proof.} 
The equations (\ref{2. 3}) and (\ref{dotW}) lead to
\begin{equation}
\begin{array}{l}
 \displaystyle 
e^{-\lambda s} \frac{d}{ds}\Bigl[ \frac12 \sum_{\alpha=1}^2 [ \rho^{(\alpha)}  \dot u^{(\alpha)}_i(s) \dot u^{(\alpha)}_i(s) +   \rho^{(\alpha)}   \K^{(\alpha)}  \dot
\varphi^{(\alpha)} (s) \dot \varphi^{(\alpha)} (s)  + 2W ({\bf U}(s)) ]\Bigr]
=
\\[5 mm] \quad
 \quad
 \displaystyle
=e^{-\lambda s}\sum_{\alpha=1}^2 \Bigl[ \rho^{(\alpha)} f^{(\alpha)}_i (s) \dot u^{(\alpha)}_i (s)  +  \rho^{(\alpha)} \ell^{(\alpha)}  (s)\dot \varphi^{(\alpha)} (s)  +
\\[5 mm] \quad  \quad
 \displaystyle
+   \bigl[ S^{(\alpha)}_{ji}({\bf U}(s)) \dot
 u^{(\alpha)}_i(s) + h^{(\alpha)}_j({\bf U}(s))  \dot \varphi^{(\alpha)} (s) \bigr]_{,j}\Bigr] . 
\end{array}
\label{A18}
\end{equation}
 By an integration of the equations (\ref{A18})
 over $P\times [0,t]$ and by using the divergence theorem, we obtain
the equation (\ref{A16}). \quad $\bull$

If we introduce
\begin{equation}
{\cal E}(t) =
\int_B  \frac12
\sum_{\alpha=1}^2
\bigl[ \rho^{(\alpha)}  \dot u^{(\alpha)}_i(t)\dot u^{(\alpha)}_i(t) +
  \rho^{(\alpha)}   \K^{(\alpha)}  \dot \varphi^{(\alpha)} (t)\dot\varphi^{(\alpha)} (t) 
 + 2W({\bf U}(t)) \bigr] dv ,
\label{energia}
\end{equation}
then, for $\lambda =0 $ and $P=B$  the equations (\ref{A16}) reduce to
\begin{equation}
\begin{array}{l}
\displaystyle 
{\cal E}(t) ={\cal E}(0) 
+\int_0^t \int_B \sum_{\alpha=1}^2 [ \rho^{(\alpha)} f^{(\alpha)}_i(s) \dot u^{(\alpha)}_i(s) +  \rho^{(\alpha)}  \ell^{(\alpha)} (s) \dot \varphi^{(\alpha)} (s)] dv ds
+
\\[5 mm] \quad \quad
\displaystyle 
+\int_0^t \int_{\partial B}  \sum_{\alpha=1}^2[  s^{(\alpha)}_i({\bf U}(s)) \dot
 u^{(\alpha)}_i(s) + h^{(\alpha)}({\bf U}(s)) \dot \varphi^{(\alpha)} (s) ] da ds  . 
\end{array}
\label{inizioenergia}
\end{equation}
We note that ${\cal E}(t) $ is a measure of the energy stored in $B$ at time $t$.

 {\bf Lemma 3.2}  Let ${\bf U}$ be a solution of 
initial-boundary-value problem ${\cal P}$. Then, for every regular region $P\subset B$ with regular boundary 
$\partial P$, it follows that 
\begin{equation}
\begin{array}{l}
\!\!
\displaystyle \int_P \sum_{\alpha=1}^2 [ \rho^{(\alpha)}  u^{(\alpha)}_i(t) \dot u^{(\alpha)}_i(t) +   \rho^{(\alpha)}   \K^{(\alpha)}  \varphi^{(\alpha)} (t) \dot
\varphi^{(\alpha)} (t) ]  dv = 
\\[5 mm] \quad
\displaystyle
= \int_0^t \int_P \sum_{\alpha=1}^2 \{ [ \rho^{(\alpha)}  
\dot u^{(\alpha)}_i(s) \dot u^{(\alpha)}_i(s) +   \rho^{(\alpha)}   \K^{(\alpha)}  \dot \varphi^{(\alpha)} (s)\dot \varphi^{(\alpha)} (s) ]
- 2 W({\bf U}(s)) \} dvds + 
\\[5 mm] \quad
\displaystyle
+\int_P \sum_{\alpha=1}^2 [ \rho^{(\alpha)}  u^{(\alpha)}_i(0) \dot u^{(\alpha)}_i(0) +
  \rho^{(\alpha)}   \K^{(\alpha)}  \varphi^{(\alpha)} (0) \dot \varphi^{(\alpha)} (0) ] dv + 
\\[5 mm] \quad
\displaystyle + \int_0^t \int_{\partial P}  \sum_{\alpha=1}^2[ s^{(\alpha)}_i({\bf U}(s)) u^{(\alpha)}_i(s) + h^{(\alpha)}({\bf U}(s))\varphi^{(\alpha)} (s)] 
da ds  +
\\[5 mm] \quad
\displaystyle
 +  \int_0^t \int_P \sum_{\alpha=1}^2 [\rho^{(\alpha)} f^{(\alpha)}_i(s) u^{(\alpha)}_i(s)  +   \rho^{(\alpha)} \ell^{(\alpha)} (s) \varphi^{(\alpha)} (s) ] dv ds
, \qquad \qquad \qquad t\in I. 
\end{array}
\label{A19}
\end{equation}

 {\bf Proof.} 
 The relations (\ref{2. 3}) and (\ref{nodotW}) imply that
\begin{equation}
\begin{array}{l}
 \displaystyle 
\frac{d}{ds}\Bigl[  \sum_{\alpha=1}^2 [ \rho^{(\alpha)}  u^{(\alpha)}_i(s) \dot u^{(\alpha)}_i(s) +   \rho^{(\alpha)}   \K^{(\alpha)} 
\varphi^{(\alpha)} (s) \dot \varphi^{(\alpha)} (s)  ]\Bigr]=
\\[5 mm] \quad
 \displaystyle
=\sum_{\alpha=1}^2 \bigl[ \rho^{(\alpha)} \dot u^{(\alpha)}_i(s) \dot u^{(\alpha)}_i(s) +   \rho^{(\alpha)}  
 \K^{(\alpha)} \dot\varphi^{(\alpha)} (s)\dot \varphi^{(\alpha)} (s) \bigr]+
\\[5 mm] \quad
 \displaystyle
+\sum_{\alpha=1}^2 \Bigl[ \rho^{(\alpha)} f^{(\alpha)}_i (s) u^{(\alpha)}_i (s)  +  \rho^{(\alpha)} \ell^{(\alpha)}  (s) \varphi^{(\alpha)} (s) - 2 W({\bf U}(s))  +
\\[5 mm] \quad
 \displaystyle
+ [  S^{(\alpha)}_{ji}({\bf U}(s)) 
 u^{(\alpha)}_i(s) + h^{(\alpha)}_j({\bf U}(s))  \varphi^{(\alpha)} (s) ]_{,j}
\Bigr] . 
\end{array}
\label{A22}
\end{equation}
The relation (\ref{A19}) follows from (\ref{A22})  by an integration of over $P\times [0,t]$ and by using the divergence theorem. \quad $\bull $

{\bf Lemma 3.3} Let ${\bf U}$ be a solution of 
initial-boundary-value problem. Then, for every regular region $P\subset B$ with regular boundary 
$\partial P$, it follows that 
\begin{equation}
\begin{array}{l}
\displaystyle
2 \int_P \sum_{\alpha=1}^2 [ \rho^{(\alpha)}  u^{(\alpha)}_i(t) \dot u^{(\alpha)}_i(t) +   \rho^{(\alpha)}   \K^{(\alpha)}  \varphi^{(\alpha)} (t) \dot \varphi^{(\alpha)} (t) ] dv 
=
\\[5 mm] \quad \displaystyle
=
 \int_P \sum_{\alpha=1}^2 [ \rho^{(\alpha)} u^{(\alpha)}_i(0)\dot u^{(\alpha)}_i(2t)
  + 
 \rho^{(\alpha)} \dot
 u^{(\alpha)}_i(0) u^{(\alpha)}_i(2t)+
\\[5 mm] \quad \displaystyle
+   \rho^{(\alpha)}   \K^{(\alpha)}  \varphi^{(\alpha)} (0) \dot \varphi^{(\alpha)} (2t) 
 + \rho^{(\alpha)}   \K^{(\alpha)}
\varphi^{(\alpha)} (2t) \dot \varphi^{(\alpha)} (0)  ] dv +  
\\[5 mm] \quad
\displaystyle  +
\int_0^t \int_P \sum_{\alpha=1}^2 [ \rho^{(\alpha)} f^{(\alpha)}_i(t-s)u^{(\alpha)}_i(t+s) - \rho^{(\alpha)} f^{(\alpha)}_i(t+s)u^{(\alpha)}_i(t-s)+
\\[5 mm] \quad
\displaystyle +
 \rho^{(\alpha)} \ell^{(\alpha)}  (t-s) \varphi^{(\alpha)} (t+s) \displaystyle -   \rho^{(\alpha)} \ell^{(\alpha)}  (t+s) \varphi^{(\alpha)} (t-s)] dv ds+
\\[5 mm] \quad
\displaystyle  + \int_0^t \int_{\partial P}
 \sum_{\alpha=1}^2 [   s^{(\alpha)}_i({\bf U}(t-s)) u^{(\alpha)}_i(t+s) -
  s^{(\alpha)}_i({\bf U}(t+s)) u^{(\alpha)}_i(t-s)+
\\[5 mm] \quad
\displaystyle + h^{(\alpha)}({\bf U}(t-s))\varphi^{(\alpha)} (t+s) - 
h^{(\alpha)}({\bf U}(t+s)) \varphi^{(\alpha)} (t-s) ] da ds , 
\qquad \quad
t\in I. 
\end{array}
\label{A23}
\end{equation}

 {\bf Proof.} 
For every function $\phi\in C^2(I)$, it is holds the following identity
\begin{equation}
\begin{array}{ll}
\displaystyle 
 \ddot
 \phi(t-s) \phi(t+s)- \ddot \phi(t+s)\phi(t-s)=&
\\[5 mm] \quad \displaystyle
  -\frac{d}{ds} \{  \dot \phi(t-s) \phi(t+s) +\phi(t-s) \dot
 \phi(t+s)   \}, 
& s\in [0,t], \quad t\in I,
\end{array}
\label{A24}
\end{equation}
In view of the relation (\ref{2. 3}), we have
\begin{equation}
\begin{array}{l}
\displaystyle
\sum_{\alpha =1}^2 [\rho^{(\alpha)}  \ddot u^{(\alpha)}_i(t-s)u^{(\alpha)}_i(t+s) -
\rho^{(\alpha)} \ddot u^{(\alpha)}_i(t+s)  u^{(\alpha)}_i(t-s)] = 
\\[5 mm]
\displaystyle
 \quad \sum_{\alpha =1}^2 \Bigl\{
\rho^{(\alpha)} f^{(\alpha)}_i(t-s)u^{(\alpha)}_i(t+s)
 - \rho^{(\alpha)} f^{(\alpha)}_i(t+s)u^{(\alpha)}_i(t-s)+
\\[5 mm] \quad
\displaystyle
+ [ S^{(\alpha)}_{ji}({\bf U}(t-s)) u^{(\alpha)}_i(t+s) 
-  S^{(\alpha)}_{ji}({\bf U}(t+s)) u^{(\alpha)}_i(t-s)]_{,j} + 
\\[5 mm] \quad
+ [ S^{(\alpha)}_{ji}({\bf U}(t+s))u^{(\alpha)}_{i,j}
(t-s) -  S^{(\alpha)}_{ji}({\bf U}(t-s))u^{(\alpha)}_{i,j}(t+s)]
\Bigr\}
+
\\[5 mm] \quad
-p_i({\bf U}(t-s)) d_i({\bf U}(t+s))+
p_i({\bf U}(t+s)) d_i({\bf U}(t-s))
,
\end{array}
 \label{A25}
\end{equation}
and
\begin{equation}
\begin{array}{l}
\displaystyle
\sum_{\alpha =1}^2 
 [ \rho^{(\alpha)} \K^{(\alpha)}
\ddot \varphi^{(\alpha)}(t-s)\varphi^{(\alpha)}(t+s) -\rho^{(\alpha)} \K^{(\alpha)} \ddot\varphi^{(\alpha)}(t+s)  \varphi^{(\alpha)}(t-s)] = 
\\[5 mm] \quad\displaystyle
 \sum_{\alpha =1}^2 \Bigl\{
\rho^{(\alpha)} \ell^{(\alpha)}(t-s)\varphi^{(\alpha)}(t+s)
 - \rho^{(\alpha)} \ell^{(\alpha)}(t+s)\varphi^{(\alpha)}(t-s)+
\\[5 mm] \quad
\displaystyle
+ [ h^{(\alpha)}_j({\bf U}(t-s)) \varphi^{(\alpha)}(t+s) 
-  h^{(\alpha)}_j({\bf U}(t+s)) \varphi^{(\alpha)}(t-s)]_{,j} + 
\\[5 mm] \quad
+ [ h^{(\alpha)}_j({\bf U}(t+s)) \varphi^{(\alpha)}_{,j}
(t-s) -  h^{\alpha)}_j({\bf U}(t-s)) \varphi^{(\alpha)}_{,j}(t+s)]
+
\\[5 mm] \quad
+g({\bf U}(t-s)) \varphi^{(\alpha)}(t+s))-
g({\bf U}(t+s)) \varphi^{(\alpha)}(t-s))\Bigr\},
\end{array}
 \label{A25bis}
\end{equation}
Further, with the help of (\ref{2. 5bis})--(\ref{2. 2bis})
we prove that
\begin{equation}
\begin{array}{l}
\displaystyle
 \sum_{\alpha =1}^2 \Bigl\{
 S^{(\alpha)}_{ji}({\bf U}(t+s))u^{(\alpha)}_{i,j}
(t-s) -  S^{(\alpha)}_{ji}({\bf U}(t-s))u^{(\alpha)}_{i,j}(t+s)
+\\[5 mm] \quad
\displaystyle
+ [ h^{(\alpha)}_j({\bf U}(t-s)) \varphi^{(\alpha)}(t+s) 
-  h^{(\alpha)}_j({\bf U}(t+s)) \varphi^{(\alpha)}(t-s)]_{,j} + 
\\[5 mm] \quad
+ [ h^{(\alpha)}_j({\bf U}(t+s)) \varphi^{(\alpha)}_{,j}
(t-s) -  h^{(\alpha)}_j({\bf U}(t-s)) \varphi^{(\alpha)}_{,j}(t+s)]
+
\\[5 mm] \quad
-p_i({\bf U}(t-s)) d_i({\bf U}(t+s))+
p_i({\bf U}(t+s)) d_i({\bf U}(t-s))+
\\[5 mm] \quad
+g({\bf U}(t-s)) \varphi^{(\alpha)}(t+s)-
g({\bf U}(t+s)) \varphi^{(\alpha)}(t-s) \Bigr\}
=0.
\end{array}
 \label{zero}
\end{equation}
Then, the equations (\ref{A24})--(\ref{zero}) imply that

\begin{equation}
\begin{array}{l}
\displaystyle
 \quad \sum_{\alpha =1}^2 \Bigl\{
\rho^{(\alpha)} f^{(\alpha)}_i(t-s)u^{(\alpha)}_i(t+s)
 - \rho^{(\alpha)} f^{(\alpha)}_i(t+s)u^{(\alpha)}_i(t-s)+
\\[5 mm] \quad
\displaystyle
+ [ S^{(\alpha)}_{ji}({\bf U}(t-s)) u^{(\alpha)}_i(t+s) 
-  S^{(\alpha)}_{ji}({\bf U}(t+s)) u^{(\alpha)}_i(t-s)]_{,j} + 
\\[5 mm] \quad\displaystyle
\rho^{(\alpha)} \ell^{(\alpha)}(t-s)\varphi^{(\alpha)}(t+s)
 - \rho^{(\alpha)} \ell^{(\alpha)}(t+s)\varphi^{(\alpha)}(t-s)+
\\[5 mm] \quad
\displaystyle
+ [ h^{(\alpha)}_j({\bf U}(t-s)) \varphi^{(\alpha)}(t+s) 
-  h^{(\alpha)}_j({\bf U}(t+s)) \varphi^{(\alpha)}(t-s)]_{,j}\Bigr\}=
\\[5 mm]
\displaystyle
\quad
=-\frac{d\,}{ds}\sum_{\alpha =1}^2
\Bigl\{ \rho^{(\alpha)} \dot u^{(\alpha)}_i(t-s)u^{(\alpha)}_i(t+s) +
\rho^{(\alpha)}u^{(\alpha)}_i(t-s)\dot u^{(\alpha)}_i(t+s)  
 \\[5 mm] \quad
\displaystyle
+\rho^{(\alpha)} \K^{(\alpha)}
 \ddot \varphi^{(\alpha)}(t-s)\varphi^{(\alpha)}(t+s) +
\rho^{(\alpha)} \K^{(\alpha)}\varphi^{(\alpha)}(t-s)
 \ddot\varphi^{(\alpha)}(t+s)  
\Bigr\} =  
\end{array}
 \label{Aderiva}
\end{equation}

The equation (\ref{A23})
 is reached by performing an integration of the equations (\ref{Aderiva}) over $P\times [0,t]$ and then  by 
using the divergence theorem.  $\bull$

 \section {A time weighted surface measure}

Having fixed a time $T\in I$, for the external given data of the
 problem ${\cal P}$ we define the set $\widehat
D_T$ by:
\newline i. \,\,\, if ${{\bf x}}{\in B},$ then
$$
a_i^{(1)} ({\bf x}) \ne 0
\hbox{\, or  \,}
\dot a_i^{(1)} ({\bf x})  \ne 0
\hbox{\, or  \,}
a_i^{(2)} ({\bf x}) \ne 0
 \hbox{\, or  \,}
\dot a_i^{(2)} ({\bf x})  \ne 0
$$
 or
$$
 \varphi^{(1)}_0 ({\bf x}) \ne 0
 \hbox{\, or  \,}
\dot \varphi^{(1)}_0 ({\bf x}) \ne 0
\hbox{\, or  \,}
\varphi^{(2)}_0 ({\bf x}) \ne 0
\hbox{\, or  \,}
\dot \varphi^{(2)}_0 ({\bf x}) \ne 0
$$
 or there exists  
$\tau\in [0,T]$ such that 
$$
f_i^{(1)}({\bf x}, \tau) \ne 0 \hbox{\, or  \,}  f_i^{(2)}({\bf x}, \tau) \ne 0
\hbox{\, or  \,}
\ell^{(1)}({\bf x}, \tau) \ne 0
 \hbox{\, or  \,}  \ell^{(2)}({\bf x}, \tau) \ne 0;$$
ii. if  ${\bf x}\in\partial B,$ then there exists $\tau\in [0,T]$  such that 
$$
s_i^{(1)}({\bf x}, \tau) \dot u_i^{(1)} ({\bf x}, \tau) \ne0
 \hbox{\, or  \,}  s_i^{(2)}({\bf x}, \tau) \dot u_i^{(2)} ({\bf x}, \tau) \ne0
$$
or 
$$
h^{(1)}({\bf x}, \tau) \dot \varphi^{(1)} ({\bf x}, \tau) \ne0
 \hbox{\, or  \,}
h^{(2)}({\bf x}, \tau) \dot \varphi^{(2)} ({\bf x}, \tau) \ne0
. $$

The set $\widehat D_T$ represents the support of the initial and boundary data and the body force on the time interval
$ [0,T]$. In what follows, we
 assume $\widehat D_T$ is a bounded set. 

We consider a nonempty set $\widehat D_T^*$ which is
such that ${\widehat D_T} \subset {\widehat D_T^*} 
\subset \bar{B} $ and  
\newline
i. \,\,\, if $ {\widehat D_T}\cap B \ne \emptyset $,
 then we choose $\widehat D_T^*$ to 
be the smallest bounded regular region in $\bar{B}$ 
that includes $\widehat D_T$ ;
 in particular, we set
 ${\widehat D_T^*} = {\widehat D_T}$ if $\widehat D_T$ 
it also happens to be a regular region;
\newline
ii. \,\, if $\emptyset \ne{\widehat D_T}\subset \partial B$,
then we choose $\widehat D_T^*$ to 
be the smallest regular subsurface of $\partial B$ 
that includes $\widehat D_T$ ;
 in particular, we set
 ${\widehat D_T^*} = {\widehat D_T}$ if $\widehat D_T$ 
is a regular subsurface of $\partial B$;
\newline
iii. \, if ${\widehat D_T} = \emptyset$, 
then we choose $\widehat D_T^*$ to  be an arbitrary nonempty regular subsurface of $\partial B$.
\bigskip

Now, we mean the set $D_r,$ by
\begin{equation}
D_r = \{ {\bf x} {\in}\bar{B} : {\widehat D_T^*} \cap{\bar\Sigma(\bf
x}, r) \ne \emptyset \}, \qquad r\geq 0,
\label{2. 14}
\end{equation}
where $\bar\Sigma({\bf x}, r)$ is the closed ball 
with radius $r$ and center
 at ${\bf x}$. Further, we use the notation $B_r$ for the part of $B$ contained in
$\bar B\setminus D_r$ and $B(r_1, r_2) = B_{r_2}
\setminus B_{r_1}, r_1>r_2$; $S_r$ denotes the 
subsurface of $\partial B_r$ contained into inside of $B$ and whose 
outward unit normal vector $\bf n$ is forwarded to the exterior of $D_r$. Of course,
taking into account that for each $r > 0$,
${\widehat D_T} \subset D_r $ and ${\widehat D_T} \cap B_r = \emptyset 
$,
we get
\begin{equation}
\begin{array}{ll}
 \tilde u_i^{(\alpha)} =0,
\quad v_i^{(\alpha)}=0, \quad
\tilde \varphi^{(\alpha)} =0,
\quad \tilde \zeta^{(\alpha)}=0 \quad &  
\hbox{ on }B_r,\quad \\[5 mm]
f_i^{(\alpha)}=0, \quad \ell^{(\alpha)}=0&
\hbox{ on }B_r\times [0,T],\\[5 mm]
s_i^{(\alpha)}\dot u_i^{(\alpha)}=0 \quad 
h^{(\alpha)}\dot \varphi^{(\alpha)}=0 \quad 
&\hbox{ on }(B_r\cap \partial B)\times [0,T]. 
\end{array}
\label{BR}
\end{equation}

For a fixed positive parameter $\lambda$ and for any $ r\ge 0$, $t\in [0,T]$,
 we define the time--weighted surface power function $P(r,t)$
\begin{equation}
P(r,t) = - \int^t_0\int_{S_r}e^{-\lambda s}
\sum_{\alpha=1}^2 \bigl[
s_i^{(\alpha)}({\bf U}(s))\dot u_i^{(\alpha)}(s )
+
h^{(\alpha)}({\bf U}(s))\dot \varphi^{(\alpha)}(s )
\bigr]\;da \, ds. \quad
\label{2. 15}\end{equation}

 In the following Lemmas, we show some relevant properties of the function $P(r,t)$.

{\bf Lemma 4.1} Let ${\bf U}$ be a solution of 
initial-boundary-value problem ${\cal P}$ and $\widehat D_T$
be the bounded support of the external data
 on the time interval $[0,T]$. Then, the corresponding 
time--weighted surface power function 
 $P(r,t)$ is a continuous differentiable function on $r\ge 0$, 
$t\in [0,T]$ and
\begin{equation}
\frac{\partial}{\partial t} P(r,t) = - \int_{S_r}
e^{-\lambda t} \sum_{\alpha=1}^2 \bigl[ 
s_i^{(\alpha)}({\bf U}(t)) \dot u_i^{(\alpha)}(t) +
h^{(\alpha)}({\bf U}(t)) \dot \varphi^{(\alpha)}(t)
 \bigr]\,da \, ;  \label{2. 18}
\end{equation} 
\begin{equation}
\begin{array}{ll}
\displaystyle\frac{\partial}{\partial r}P(r,t) = 
&
\displaystyle
- \frac12 \int_{S_r}
e^{-\lambda t} \sum_{\alpha=1}^2 \Bigl[
\rho^{(\alpha)} \dot u_i^{(\alpha)}(t) \dot u_i^{(\alpha)}(t) +
\rho^{(\alpha)}  \K^{(\alpha)}  \dot \varphi^{(\alpha)}(t) \dot
\varphi^{(\alpha)}(t) +
\\[5 mm]
& +
 2 W({\bf U}(t))\Bigr] \; da -
 \displaystyle\frac{\lambda}{2}
\int^t_0\int_{S_r}e^{-\lambda s}\sum_{\alpha=1}^2
\Bigl[\rho^{(\alpha)} \dot
u_i^{(\alpha)}(s) \dot u_i^{(\alpha)}(s) +
\\[5 mm]
& +
\rho^{(\alpha)}  \K^{(\alpha)}  \dot \varphi^{(\alpha)}(s) \dot
\varphi^{(\alpha)}(s)+ 2 W({\bf U}(s))
\Bigr]\, da \, ds. \\[5 mm]
\end{array}
\label{2. 17}
\end{equation} 

Moreover, at a fixed  $t\in [0,T]$, $P(r,t)$ is a non--increasing
function with respect to $r$. 
\smallskip

{ \bf Proof. }
 The equation (\ref{2. 18}) is an immediate consequence of the definition of
$P(r,t)$. 
\smallskip

The Lemma 3.1 for $B(r_1,r_2)$ 
 and the divergence theorem imply
\smallskip
\begin{equation}
\begin{array}{l}
P(r_1,t) - P(r_2,t) = 
\\[5 mm]
\qquad \displaystyle
 =\int^t_0\int_{B(r_1,r_2)}e^{-\lambda s} 
\sum_{\alpha=1}^2
\bigl[\rho^{(\alpha)}f_i^{(\alpha)}(s) \dot u_i^{(\alpha)}(s) 
+\rho^{(\alpha)} \K^{(\alpha)}\ell^{(\alpha)}(s) \dot \varphi^{(\alpha)}(s) 
\bigr]\;
dv ds +
\\[5 mm]
\qquad - \displaystyle
\frac12\int^t_0\int_{B(r_1,r_2)}e^{-\lambda s} \frac{\partial}{
\partial s}
 \sum_{\alpha=1}^2
\bigl[ \rho^{(\alpha)} \dot u_i^{(\alpha)}(s) \dot u_i^{(\alpha)}(s) 
+\rho^{(\alpha)} \K^{(\alpha)}\dot\varphi^{(\alpha)}(s) \dot
\varphi^{(\alpha)}(s) +
\\[5 mm]
\qquad 
+ 2W({\bf U})(s))
 \bigr]\;dv ds, 
\end{array}
\label{2. 25}
\end{equation}
with $0 \le r_2\le r_1 $ and fixed $t>0$. 
By using equations
 (\ref{BR}) and performing an integration by parts,
the relation (\ref{2. 25}) becomes
\begin{equation}
\begin{array}{l}
P(r_1,t) - P(r_2,t) =-\displaystyle\frac12\int_{B(r_1,r_2)}
e^{-\lambda t}
\sum_{\alpha=1}^2 
\bigl[\rho^{(\alpha)} \dot u_i^{(\alpha)}(t)\dot u_i^{(\alpha)}(t) +
\rho^{(\alpha)} \K^{(\alpha)}\dot\varphi^{(\alpha)}(t) \dot
\varphi^{(\alpha)}(t)+ 
\\[5 mm]\qquad \displaystyle
+ 2 W({\bf U}(t))\bigr]\,dv 
- \frac{\lambda}{2}\int^t_0\int_{B(r_1,r_2)}e^{-\lambda s}
\sum_{\alpha=1}^2 \bigl[\rho^{(\alpha)} 
\dot u_i^{(\alpha)}(s) \dot u_i^{(\alpha)}(s) +
\rho^{(\alpha)} \K^{(\alpha)}\dot\varphi^{(\alpha)}(s) \dot
\varphi^{(\alpha)}(s)+ 
\\[5 mm]
+2 W({\bf U}(s))\bigr]\,dv \, ds. 
\end{array}
 \label{2. 16}
\end{equation}
This relation straighty leads to (\ref{2. 17}).

Since we have assumed that $\rho^{(\alpha)}$,  $ \K^{(\alpha)}$ are
strictly positive, $\lambda$ is positive and $W({\bf U})$ is a positive 
definite quadratic form, the equation (\ref{2. 16}) gives
\begin{equation}
P(r_1,t) \leq P(r_2,t) \qquad \qquad
\hbox{with } r_1
\ge r_2. \qquad \qquad \bull
 \label{positiva}
\end{equation}
\smallskip

{\bf Lemma 4.2} Let ${\bf U}$ be a solution of 
 initial-boundary-value problem ${\cal P}$  and $\widehat D_T$ be the bounded support
of the external data
 on the time interval $[0,T]$. Then, the function $P(r,t)$ satisfies the following first--order differential inequalities, for any $r\ge 0$, 
$t\in [0,T]$  
\begin{equation} 
\frac{\lambda}{c}\left|P(r,t)\right| + \frac{\partial}
{\partial r}P(r,t) \le 0,  
\label{2. 19}
\end{equation} 
\begin{equation} \frac{1}{c}\left|\frac{\partial}{\partial t}P(r,t)\right| +
\frac{\partial}{\partial r}P(r,t) \le 0,  
\label{2. 20}
\end{equation} 
where 
\begin{equation} 
c = \displaystyle\sqrt{\frac{\xi_M  }{ m}} 
\qquad \quad \hbox{ with }\; m=\hbox{min }\{\rho^{(1)},\rho^{(2)}, 
\rho^{(1)} \K^{(1)},\rho^{(2)} \K^{(2)} \}. 
 \label{2. 21}
\end{equation}
and $\xi_M $ is the maximum elastic moduli.

{\bf Proof. } It follows from Schwarz's inequality and the
arithmetic--geometric mean inequality
\begin{equation}
\begin{array}{l}
\displaystyle
\left|\sum_{\alpha=1}^2 \Bigl[
s_i^{(\alpha)}({\bf U}(t))\dot u_i^{(\alpha)}(t)+
h^{(\alpha)}({\bf U}(t))\dot \varphi^{(\alpha)}(t)
\Bigr]\right|
\le 
\\[5 mm]
\displaystyle
\le\frac 12
\sum_{\alpha=1}^2 \Bigl[ \varepsilon  \rho^{(\alpha)} \dot u_i^{(\alpha)}(t)
\dot u_i^{(\alpha)}(t)+
\frac{1}{\varepsilon \rho^{(\alpha)}}
 s_i^{(\alpha)}({\bf U}(t))s_i^{(\alpha)}({\bf U}(t)) +
\\[5 mm]
\displaystyle
+
\varepsilon  \rho^{(\alpha)} \K^{(\alpha)} \dot \varphi^{(\alpha)}(t)
\dot \varphi^{(\alpha)}(t)+
\frac{1}{\varepsilon \rho^{(\alpha)}  \K^{(\alpha)}} h^{(\alpha)}({\bf
U}(t))h^{(\alpha)}({\bf U}(t))\Bigr],
\end{array}
\label{Schw}
\end{equation}
where $\varepsilon$ is an arbitrary positive constant. 

Using the relations  
(\ref{2. 15}, \ref{Schw}, \ref{2. 21}, \ref{okok})
and taking $\varepsilon =c$, 
 we deduce 
\begin{equation}
\begin{array}{ll}
\left|P(r,t)\right| \le \displaystyle\frac{c}{2}
\int^t_0\int_{S_r}   
& \displaystyle
e^{-\lambda s}
 \sum_{\alpha=1}^2
\bigl[\rho^{(\alpha)} \dot u_i^{(\alpha)}(s) \dot u_i^{(\alpha)}(s)
+\rho^{(\alpha)}  \K^{(\alpha)} \dot \varphi^{(\alpha)}(s) \dot
\varphi^{(\alpha)}(s)+
\\[5 mm]
&
+2 W({\bf U}(s)) \bigr]\,da \, ds \, 
\end{array}
\label{2. 27}
\end{equation}
for $r\ge0$ and $ 0 \le t \le T$. 

Similarly, from (\ref{2. 18}) we obtain
\begin{equation}
\begin{array}{ll}
\displaystyle
\left|\frac{\partial}{\partial t} P(r,t)\right| \le \displaystyle\frac{c}{2}
\int_{S_r}
&
\displaystyle
e^{-\lambda t}
 \sum_{\alpha=1}^2
\bigl[\rho^{(\alpha)} \dot u_i^{(\alpha)}(t) \dot u_i^{(\alpha)}(t) 
+\rho^{(\alpha)}  \K^{(\alpha)} \dot \varphi^{(\alpha)}(t) \dot
\varphi^{(\alpha)}(t)+
\\[5 mm]
&
+ 2 W({\bf U}(t)) \bigr]\,da \,. 
\end{array}
\label{2. 27bis}
\end{equation}
for $r\ge0$ and $t\in [0,T]$.
By the equation (\ref{2. 17}) and 
the relations (\ref{2. 27}) and (\ref{2. 27bis}) 
we obtain the result. $ \bull$
\medskip

{\bf Lemma 4.3}  Let ${\bf U}$ be a solution of 
initial-boundary-value problem ${\cal P}$ and $\widehat D_T$ be the
bounded support of the external data
 on the time interval $[0,T]$. Then, it follows that
\smallskip
\begin{equation}
P(r,t) \ge 0, \quad \hbox{ for } r\ge 0, \quad 0 \le t \le T \, ;
 \label{2. 22}
\end{equation}
moreover 
\begin{equation}P(r,t) = {\it E} (r,t) \,,
 \label{2. 23}\end{equation}
\medskip
where
\smallskip
\begin{equation}
\begin{array}{ll}\!
\displaystyle 
 E (r,t) = 
\frac12\int_{B_r}
&
\displaystyle
e^{-\lambda t}\sum_{\alpha=1}^2
\Bigl[\rho^{(\alpha)} \dot u_i^{(\alpha)}(t) \dot u_i^{(\alpha)}(t)
+\rho^{(\alpha)}  \K^{(\alpha)}\dot \varphi^{(\alpha)}(t) \dot
\varphi^{(\alpha)}(t)
+ 
\\[5 mm]
&+2 W({\bf U}(t))\Bigr]\;dv
 + \displaystyle\frac{\lambda}{2}\int^t_0\int_{B_r}e^{-\lambda
s}\sum_{\alpha=1}^2 \Bigl[\rho^{(\alpha)} \dot u_i^{(\alpha)}(s) \dot
u_i^{(\alpha)}(s) 
+ 
\\[5 mm]
&
+\rho^{(\alpha)}  \K^{(\alpha)}\dot \varphi^{(\alpha)}(s) \dot
\varphi^{(\alpha)}(s)
+ 2 W({\bf U}(s))\Bigr]\; dv \, ds \,. 
\end{array}
\label{2. 24}
\end{equation}

{\bf Proof. } If B is a bounded
body, then the variable $r$ ranges on $[0,L]$, where
\begin{equation}
L = {max} \, \{ {min} \{ \, [(x_i - y_i)(x_i - y_i)]^{\frac12} :
{\bf y} \in {\widehat D_T^*} \} : {\bf x} \in \bar B \, \} <\infty. 
 \label{2. 28}\end{equation}

Starting from the definition of $\widehat D_T$ and 
by using the relation (\ref{2. 15}), we obtain
\begin{equation}
P(L,t) = 0, \quad 0 \le t \le T \, ;
 \label{2. 29}\end{equation}
thus, the equation (\ref{positiva}) implies the relation
 (\ref{2. 22}). 
\smallskip

If B is an unbounded
body, then the variable $r$ ranges on 
$[0,\infty)$. The inequality (\ref{2. 20}) is equivalent to 
\begin{equation} \displaystyle\frac{1}{c} \,
 \frac{\partial}{\partial t}P(r,t) + \frac
{\partial}{\partial r}P(r,t) \le 0, 
\label{2. 30}\end{equation}
\smallskip
and
\smallskip
\begin{equation} \displaystyle - \frac{1}{c} \, \frac{\partial}{\partial t}P(r,t) + \
\frac{\partial}{\partial r}P(r,t) \le 0. 
\label{2. 31}
\end{equation} \smallskip

If we choose the initial condition $(r_0, t_0)$ such that
$t_0\in[0,T]$ and $r_0 \ge c t_0$  and we put
$\displaystyle t=t_0+\frac{ r -r_0}{c}$ 
in inequality
(\ref{2. 30}), then 
 \smallskip \begin{equation}
\displaystyle\frac{d}{dr} \, [P(r, t_0 + \frac{r - r_0}{c}
) \, ]  \le 0, \label{2. 32}
\end{equation} 
thus,
\begin{equation}
P(r, t_0 + \frac{r - r_0}{c} ) \, 
\le P(r_1, t_0 + \frac{r_1 - r_0}{c} ) \, \qquad\hbox{ with }
r\ge r_1. 
 \label{2. 32bis}
\end{equation}
For $r=r_0$ and $r_1 = r_0 - ct_0$, we get
\begin{equation}P(r_0,t_0) \le P(r_0 - c t_0, 0). \label{2. 33}\end{equation}

Similarly, by setting $\displaystyle t=t_0-\frac{ r
-r_0}{c}$ in (\ref{2. 31}), it follows
\smallskip
\begin{equation} \displaystyle\frac{d}{dr} \, [P(r, t_0 - \frac {r - r_0}{c} ) \, ] \le \
0, \label{2. 34}\end{equation}
so that
\begin{equation}P(r_0 + c t_0, 0) \le P(r_0,t_0). 
 \label{2. 35}\end{equation}
Taking into account 
$P(r_0 - c t_0, 0)=0$ and $P(r_0 + c t_0, 0)=0$, 
the relations (\ref{2. 33}), (\ref{2. 35}) imply that
$$
P(r_0,t_0) = 0. 
$$
Of course, for $r_0\to \infty$ in the above relations, it follows
\smallskip
\begin{equation}
P(\infty, t_0) = \lim_{r_0 \to \infty} P(r_0,t_0) = 0,
 \label{2. 36}\end{equation}
\smallskip
and, by (\ref{positiva}), we conclude that the relation
 (\ref{2. 22}) is true. 

The equation (\ref{2. 23}) follows from the relation (\ref{2. 16}) by means of the use of the
relations (\ref{2. 29}) and (\ref{2. 36}). \quad 
 $\bull$
\medskip

\section{Spatial behaviour}

By the properties of the time-wieghted surface power function $P$, we establish  the theorem that gives a complete description of the spatial behaviour of the elastic process in question outside of the support of the external data.

{\bf Theorem 5.1}
 Let ${\bf U}$ be a solution of 
initial-boundary-value problem, $\widehat D_T$ be the bounded support
of the external data
 on the time interval $[0,T]$ and let $P(r,t)$ be
the time--weighted surface power
 measure associated with ${\bf U}$. 

i. {\it Spatial behaviour: } For 
each fixed $t\in[0,T]$ and $0 \le r \le c t$, we have
\begin{equation}
P(r,t) \le P(0,t) \, {exp} \,
 (- \displaystyle\frac{\lambda}{c} r ) 
;
\label{2. 38}
\end{equation}

ii. {\it Domain of influence results:} For 
each fixed $t\in[0,T]$ and $r\ge c t$, we have
\begin{equation}
\begin{array}{ll}
 u_i^{(1)}= 0, \quad  u_i^{(2)} = 0,
&
\\[5 mm]
\varphi^{(1)}= 0, \quad  \varphi^{(2)} = 0
\qquad \qquad\hbox{on } B_{r}\times [0, T].
\end{array}
\label{2. 37}
\end{equation}

{\bf Proof. } The equations
 (\ref{2. 19}), (\ref{2. 22}) give
\begin{equation}
\frac{\partial}{\partial r} \, 
[ {exp}({\frac{\lambda}{c} r}) \, 
P(r,t) \, ] \le 0, \quad 0\le r\le {c t }, \quad 
0\le t
\le T \,, \label{2. 43}
\end{equation}
so that we obtain the equation (\ref{2. 38}).

 If we choose $t\in [0, T]$ and 
we set $r = c t $ in (\ref{2. 30}), then
\smallskip
\begin{equation} \displaystyle\frac{d}{dt} \, [ P(c t, t) ] 
\le 0 \,, \label{2. 39}\end{equation}
\smallskip
and 
\smallskip
\begin{equation}
P(r^\prime,t) \le P(c t, t) \le P(0,0) = 0, \quad \qquad \forall \, r^\prime\ge c t. 
 \label{2. 40}
\end{equation}
In order to (\ref{2. 40}), (\ref{positiva}), we have for all
$r\ge c t$
 \begin{equation}
P(r,t)  \le  0, \quad \qquad \  
 \label{2. 40tris}
\end{equation}
and, taking into account  (\ref{2. 22}),
 we obtain
 \begin{equation}
P(r,t)= 0. 
 \label{2. 40bis}
\end{equation}
 
Now, the equations (\ref{2. 40bis}), (\ref{2. 23}), (\ref{2. 24})
imply \begin{equation}
\begin{array}{ll}
 E (r,t) = 
&\displaystyle\frac12\int_{B_r}e^{-\lambda t}\sum_{\alpha=1}^2
\Bigl[\rho^{(\alpha)} \dot u_i^{(\alpha)}(t) \dot u_i^{(\alpha)}(t)
+\rho^{(\alpha)}  \K^{(\alpha)}\dot \varphi^{(\alpha)}(t) \dot
\varphi^{(\alpha)}(t)
+ 
\\[5 mm]
&+2 W({\bf U}(t))\Bigr]\;dv
 + \displaystyle\frac{\lambda}{2}\int^t_0\int_{B_r}e^{-\lambda
s}\sum_{\alpha=1}^2 \Bigl[\rho^{(\alpha)} \dot u_i^{(\alpha)}(s) \dot
u_i^{(\alpha)}(s) 
+ 
\\[5 mm]
&
+\rho^{(\alpha)}  \K^{(\alpha)}\dot \varphi^{(\alpha)}(s) \dot
\varphi^{(\alpha)}(s)
+ 2 W({\bf U}(s))\Bigr]\; dv \, ds \, =0. 
\end{array}
\label{2. 24bis}
\end{equation}
Since $\rho^{(\alpha)}$ and   $ \K^{(\alpha)}$ are strictly positive,
$\lambda$ is positive
and $W({\bf U})$ is a positive definite quadratic form, 
we  have
\begin{equation}
\dot u_i^{(1)}= 0, \quad \dot u_i^{(2)} = 0,
\quad
\dot \varphi^{(1)}= 0, \quad \dot \varphi^{(2)} = 0
\qquad \qquad\hbox{on } B_{r}\times [0, T] ;
\label{dominiobis}
\end{equation}
so, by (\ref{BR}), 
we obtain the equations (\ref{2. 37}). \quad $\bull$

If  we put $t=T$ and $r= c T$  in
the equations (\ref{2. 37}) for $t=T$ and $r= c T$ they imply
that the set $D_{cT}$ covers a domain  of elastic
disturbances produced by the the data at time $T$, i.e. 
\begin{equation}
u_i^{(1)} = 0, \quad u_i^{(2)} = 0
\quad 
\varphi^{(1)} = 0, \quad \varphi^{(2)} = 0
\qquad \qquad \hbox{ on }  B_{cT}\times [0, T].
\label{dominio}
\end{equation} 
This result is known as a so-called domain of influence
theorem (see Gurtin \cite{6})

As an immediata consequence of the  equation (\ref{2. 37}),  
we  establish the following 
 uniqueness result valid for a bounded or unbounded body

{\bf Theorem 5.2}({\it Uniqueness})  It exists at most
one (regular) solution  for the boundary-initial-value
problem. \smallskip

{ \bf Proof. } Thanks to the linearity of the problem,
 we have only to show that the null data imply null  
solution. Let $\tilde {\bf U}=\{\tilde {\bf u}^{(1)},
\tilde {\bf u}^{(2)}, \tilde
\varphi^{(1)},\tilde\varphi^{(2)}\}$ a solution
corresponding to null data. Since the set $\widehat
D_T=\emptyset  $  for each $T\in (0, +\infty)$ and the
function $P(r,t)=0$, we can conclude that  
$$ \tilde
u_i^{(1)}=0, \qquad \tilde u_i^{(2)}=0, \quad  \tilde
\varphi^{(1)} = 0, \quad \tilde\varphi^{(2)} = 0 \qquad
\hbox{ on } B\times I. \qquad \bull $$

We remark that if $B$ is a bounded regular region, 
 for values of T big enough, then it exists a value  of
$t\in [0,T]$ having the property that $D_{ct} = B$, 
the relation (\ref{2. 37}) becomes superfluous and  the
behaviour of  solutions is full described  by the relation
(\ref{2. 38}). On the other hand, for values of $T$
sufficiently small, the behaviour of solutions will  be
described  by the relation (\ref{2. 37})  almost as in B.
Similar arguments are valid for an unbounded regular
region.

 \section{Asymptotic equipartition of energy} 

 Throughout this section we study the time asympototic
behaviour of the solutions of 
the initial-boundary value problem ${\cal P}_0$ for the bounded regular
region $B$
defined by
the following equations of motion 
\begin{equation}
\begin{array}{l}
S_{ji,j}^{(\alpha)} +( -1 )^{\alpha} p_i = \rho^{(\alpha)} \ddot
{u}_i^{(\alpha)}, \\[5 mm]
h_{i,i}^{(\alpha)} +g^{(\alpha)} = \rho^{(\alpha)}  \K^{(\alpha)}\ddot
{\varphi}^{(\alpha)},
\qquad \qquad \hbox{ on }B\times (0, +\infty),
\end{array}
\label{zero2. 3}
\end{equation}
 the geometrical
equations (\ref{geom}) and the constitutive equations 
 (\ref{2. 2}), the initial conditions (\ref{inizio})
 and the  boundary conditions
\begin{equation}
\begin{array}{l}
u_i^{(\alpha)}=0\quad \hbox{ on } \bar \Sigma_1 \times I,
\qquad 
 s_i^{(\alpha)}=0
\quad \hbox{ on }  \Sigma_2 \times I ,
\\[5 mm]
\varphi^{(\alpha)}= 0,\quad \hbox{ on } \bar \Sigma_3 \times I,
\qquad
h^{(\alpha)}= 0\quad \hbox{ on }  \Sigma_4 \times I.
\end{array}
\label{zerobordo}
\end{equation}

Now,  we introduce the
Ces{\` a}ro means of various energies associated with the solution
${\bf U}$ of the problem ${\cal P}_0$ :
\begin{equation}
\begin{array}{l}
 \displaystyle
{\cal K}_{C}^u(t) = 
 \sum_{\alpha=1}^2
\frac{1}{2t} 
\int_0^t\int_B
\rho^{(\alpha)} \dot u^{(\alpha)}_i(s) \dot u^{(\alpha)}_i(s)dv ds, 
\\[5 mm]  \displaystyle 
{\cal K}_{C}^{\varphi}(t) =  \sum_{\alpha=1}^2
\frac{1}{2t} \int_0^t\int_B   \rho^{(\alpha)}  \K^{(\alpha)} 
\dot \varphi^{(\alpha)} (s)\varphi^{(\alpha)} (s) dv ds , 
\\[5 mm]  \displaystyle 
S_{C}(t) = \sum_{\alpha=1}^2\frac{1}{t} \int_0^t\int_B  W({\bf U}(s)) dv ds ,
\end{array}
 \label{A48}
\end{equation}
and 
\begin{equation}
{\cal K}_{C}(t) = {\cal K}_{C}^u(t) + 
{\cal K}_{C}^{\varphi}(t), 
\end{equation}

If $ meas \Sigma_1  = 0$ , then there exists a
family of rigid motions and null change in volume fraction which satisfy the
equations (\ref{zero2. 3}, \ref{geom}, \ref{2. 2}, \ref{inizio})
and (\ref{zerobordo}).
We decompose the initial data $a_i^{(\alpha)}$ and
$\dot a_i^{(\alpha)}$ as
\begin{equation}
a_i^{(\alpha)} = \bar a_i^{(\alpha)} + A_i^{(\alpha)}
,\quad  \quad 
\dot a_i^{(\alpha)} = \dot {\bar a_i}^{(\alpha)} + \dot A_i^{(\alpha)}
 , \label{A49}
\end{equation}
where $\bar a_i^{(\alpha)}$ and $\dot {\bar a_i}^{(\alpha)}$
 are the rigid displacements determined so that
$ A_i^{(\alpha)}$ and $\dot  A_i^{(\alpha)}$ satisfy the normalization restrictions
\begin{equation}
\begin{array}{l}
\displaystyle
 \int_B \rho^{(\alpha)}  A_i^{(\alpha)} dv = 0 ,\qquad \qquad \int_B
 \rho^{(\alpha)} \varepsilon_{ijk}\,x_j  A_k^{(\alpha)}dv = 0 ,
\\[5 mm] \displaystyle
\int_B \rho^{(\alpha)} \dot A_i^{(\alpha)} dv = 0 ,\qquad \qquad \int_B
 \rho^{(\alpha)} \varepsilon_{ijk}\,x_j \dot A_k^{(\alpha)}dv = 0 , 
\end{array}
\label{A50}
\end{equation}
and $\varepsilon_{ijk}$ is the alternating symbol.

We put
$$
\begin{array}{ll}
{\hat {\bf C}}^1(B) \equiv 
&\{ {\bf v} \hbox{ whit } v_i \in C^1(
{\bar B}) : \;  \; v_i = 0 \;\hbox{on } \; \Sigma_1
\hbox{ if } meas  \Sigma_1 \ne 0 
 ,
\\[5 mm]
&\displaystyle
 \hbox{ or } \int_B \sum_{\alpha=1}^2 \rho^{(\alpha)} v_i dv = 0 \quad ,
\quad \int_B \sum_{\alpha=1}^2 \rho^{(\alpha)} \varepsilon_{ijk}x_jv_k dv = 0 \}
\hbox{  if } meas  \Sigma_1 = 0 ,
 \end{array}
$$
and
$$
{\hat C}^1(B) \equiv \{ \zeta \in C^1(B) :\quad  \zeta = 0 \hbox{ on }
 \Sigma_3 \} ,
$$
and 
$$
{\hat {\bf W}}_1(B) \equiv \hbox{the completion of} \quad {\hat {\bf C}}^1(B)
\quad \hbox{by means of } \quad \| \cdot \|_{{\bf W}_1(B)} ,$$
and 
$$
{\hat W}_1(B) \equiv \hbox{the completion of } \quad {\hat C}^1(B) \
\hbox{by means of } \quad \| \cdot \|_{W_1(B)}.
$$
The spaces  $W_m(B)$ represents the familiar Sobolev space
and ${\bf W}_m(B) \equiv [W_m(B)]^3$.

The equation (\ref{ste}) assures that the following Korn's inequality
\cite{Korn} holds 
\begin{equation}
\displaystyle \int_B \sum_{\alpha=1}^2 2 W({\bf V}) dv \ge m_1 
\int_B \sum_{\alpha=1}^2 (v_i^{(\alpha)} v_i^{(\alpha)} + \K^{(\alpha)}
\phi^{(\alpha)}\phi^{(\alpha)} ) dv , \qquad m_1 = \hbox {const.} > 0 ,
\label{A51}
\end{equation}
for every $ {\bf V} = \{{\bf v}^{(1)}, {\bf v}^{(2)}, \phi^{(1)}, \phi^{(2)} \} : \quad
 {\bf v}^{(\alpha)} \in {\hat {\bf W}}_1(B) ,
\quad \phi^{(\alpha)} \in {\hat W}_1 (B)$.

If {\it meas}$\Sigma_1 = 0 $ , then we shall find it a convenient
practice to decompose the solution
${\bf U} = \{ {\bf u}^{(1)}, {\bf u}^{(2)}, \varphi^{(1)}, \varphi^{(2)} \}$
of the problem ${\cal P}_0$ in the form
\begin{equation}
 u_i^{(\alpha)} = \bar a_i^{(\alpha)} + t \dot {\bar a}_i^{(\alpha)} +
 v_i^{(\alpha)} \quad , \quad
 \varphi^{(\alpha)} = \phi^{(\alpha)},
 \label{A52}
\end{equation}
where ${\bf V} = = \{{\bf v}^{(1)}, {\bf v}^{(2)}, \phi^{(1)}, \phi^{(2)} \} \in {\hat {\bf W}}_1(B)\times {\hat {\bf W}}_1(B)
\times {\hat  W}_1(B) \times {\hat W}_1(B)$
 represents the solution of the
problem ${\cal P}_0$ with the initial data $\{ \bar a_i^{(\alpha)},
 \varphi_0^{(\alpha)}\}$ and $\{ \dot {\bar a}_i^{(\alpha)},
\dot \varphi_0^{(\alpha)}\}$.

 {\bf Theorem 6.1}  Let ${\bf U} $
is the solution to the problem ${\cal P}_0$. Then, for all choise of initial data
with ${\bf a}^{(\alpha)} \in {\bf W}_1(B)$ , $\dot {\bf a}^{(\alpha)} \in {\bf
W}_0(B)$,
 $ \varphi^{(\alpha)}_0 \in W_1(B)$ ,
$\dot \varphi^{(\alpha)}_0 \in W_0(B)$. 
Then the following asymptotic behaviour
of the solution ${\bf U}$ holds:

i.\, if \, ${\it meas} \Sigma_1  \ne 0$ , we have
\begin{equation}
\lim_{t \to \infty} {\cal K}_{C}(t) = \lim_{t \to \infty} S_{C}(t) =
\frac12 {\cal E}(0) ;
\label{A53}
\end{equation}

ii. if \, ${\it meas} \Sigma_1  = 0$, we have
\begin{equation}
\begin{array}{l}
\displaystyle
 \lim_{t \to \infty} {\cal K}_{C}(t) = \lim_{t \to \infty}
S_{C}(t) + \frac12 \int_B \sum_{\alpha=1}^2 \rho^{(\alpha)} 
\dot a_i^{(\alpha)} \dot a_i^{(\alpha)} dv = 
\\[5 mm]
\displaystyle
\quad=\frac12
{\cal E}(0) +
\frac12 \int_B \sum_{\alpha=1}^2 \rho^{(\alpha)} \dot a_i^{(\alpha)}
 \dot a_i^{(\alpha)} dv ,
\end{array}
\label{A54}
\end{equation}
where ${\cal E}(t)$ is defined by (\ref{energia}).

 {\bf Proof.} 
Respect to the problem ${\cal P}_0$ it follows that  ${\bf f}^{(1)}={\bf 0}$, ${\bf f}^{(2)}={\bf 0}$, $\ell^{(1)}=0$, $\ell^{(2)}=0$ 
and  the boundary conditions  (\ref{zerobordo}) are verified. Thus, the equation (\ref{inizioenergia}) becomes
\begin{equation}
{\cal E}(t) = {\cal E}(0) , \quad t\ge 0 . 
\label{A57}
\end{equation}
so that
\begin{equation}
{\cal K}_C (t) + S_C (t)  = {\cal E}(0) , \quad 
\hbox{for all} \quad t \geq 0 . \label{A58}
\end{equation}

On the other hand, the relations (\ref{A19}) and (\ref{A23}) imply 
\begin{equation}
\begin{array}{l}
\displaystyle {\cal K}_C (t) - S_C (t) = - \frac{1}{4t} \int_B \sum_{\alpha=1}^2 
\Bigl\{
2\rho^{(\alpha)} u^{(\alpha)}_i(0)\dot u^{(\alpha)}_i(0) + 
 2 \rho^{(\alpha)}   \K^{(\alpha)}  \varphi^{(\alpha)} (0) \dot \varphi^{(\alpha)} (0)  +
\\[5 mm] \quad\displaystyle
+ \rho^{(\alpha)} \{ u^{(\alpha)}_i(0) \dot u^{(\alpha)}_i(2t)
+  \dot u^{(\alpha)}_i(0) u^{(\alpha)}_i(2t) +  
 \rho^{(\alpha)}   \K^{(\alpha)}   \varphi^{(\alpha)} (0) \dot \varphi^{(\alpha)} (2t) +
\\[5 mm] \quad\displaystyle
+
 \rho^{(\alpha)}   \K^{(\alpha)} \varphi^{(\alpha)} (2t) \dot \varphi^{(\alpha)} (0) \Bigr\} dv , \qquad \quad t> 0.
\end{array}
 \label{A59}
\end{equation}
The relations (\ref{ste}), (\ref{energia}) and (\ref{A57}) imply
\begin{equation}
\begin{array}{l}  \quad
\displaystyle \int_B  \rho^{(\alpha)} \dot u^{(\alpha)}_i(s) \dot u^{(\alpha)}_i(s) dv \le 2 {\cal E}(0) ,
\\[5 mm] \quad
\displaystyle \int_B  \rho^{(\alpha)}   \K^{(\alpha)} \dot \varphi^{(\alpha)} (s)\varphi^{(\alpha)} (s) dv \le 2 {\cal E}(0) ,
\\[5 mm] \quad
\displaystyle \int_B  \varphi^{(\alpha)} (s)\varphi^{(\alpha)} (s) dv \le \frac{2}{
\xi_m} \int_B \sum_{\alpha=1}^2 W({\bf U}(s))
dv \le \frac{2}{\xi_m} {\cal E}(0) , 
\end{array}
\label{A60}
\end{equation}
thus
\begin{equation}
\begin{array}{l}
\displaystyle 
\lim_{s \to \infty} \frac 1s\int_B  \rho^{(\alpha)} \dot u^{(\alpha)}_i(s) \dot u^{(\alpha)}_i(s) dv = 0,
\\[5 mm] \displaystyle 
\lim_{s \to \infty} \frac 1s\int_B  \rho^{(\alpha)}   \K^{(\alpha)} \dot \varphi^{(\alpha)} (s)\varphi^{(\alpha)} (s) dv = 0 ,
\\[5 mm] \displaystyle 
\lim_{s \to \infty} \frac 1s\int_B  \varphi^{(\alpha)} (s)\varphi^{(\alpha)} (s) dv=0. 
\end{array}
\label{A60bis}
\end{equation}
By using the Schwarz's inequality and the relation (\ref{A60bis}) in (\ref{A59}), we obtain
\begin{equation}
\displaystyle \lim_{t \to \infty}{\cal K}_C (t) - \lim_{t \to \infty}
S_C (t) = \lim_{t \to \infty} \frac{1}{4t} 
\int_B \sum_{\alpha=1}^2 \rho^{(\alpha)} \dot u^{(\alpha)}_i(0)
 u^{(\alpha)}_i(2t) dv  .
 \label{A61}
\end{equation}

When ${\it meas} \Sigma_1 \ne 0$, then for ${\bf u}\in
{\hat {\bf W}}_1(B)$ , $\varphi^{(\alpha)} \in {\hat W}_1(B)$ , the 
relations (\ref{energia}), (\ref{A51}) and (\ref{A57}) imply
\begin{equation}
 \int_B \sum_{\alpha=1}^2 u^{(\alpha)}_i(s) u^{(\alpha)}_i(s) dv \le \frac{1}{m_1} \int_B \sum_{\alpha=1}^2 2 W({\bf U}
(s)) dv \le \frac{2}{m_1} {\cal E}(0) ,
\end{equation}
and, by means of the Schwarz's inequality, we obtain
\begin{equation}
\displaystyle \lim_{t \to \infty} \{ \frac{1}{4t} \int_B 
\sum_{\alpha=1}^2 \rho^{(\alpha)} \dot u^{(\alpha)}_i(0)
 u^{(\alpha)}_i(2t) dv \} = 0 . \label{A62}
\end{equation}
Then, by the equations (\ref{A61}) and (\ref{A62}) we have
\begin{equation}
\lim_{t \to \infty} {\cal K}_C (t) - \lim_{t \to \infty} S_C (t) = 0 .
\label{kc}
\end{equation}
The relations (\ref{A58}) and (\ref{kc}) imply 
 (\ref{A53}).
\smallskip

When ${\it meas} \Sigma_1 = 0$, then,
the equation  (\ref{A49}), (\ref{A50}) and (\ref{A52}) lead to 
\begin{equation}
\begin{array}{l}
\displaystyle \frac{1}{4t} \int_B \sum_{\alpha=1}^2 \rho^{(\alpha)} \dot u^{(\alpha)}_i(0) u^{(\alpha)}_i(2t) dv =
\frac{1}{4t} \int_B \sum_{\alpha=1}^2 \rho^{(\alpha)} \dot {\bar a}_i {\bar a}_i dv +
\\[5 mm]\displaystyle 
+ \frac{1}{4t} \int_B \sum_{\alpha=1}^2 \rho^{(\alpha)}
(\dot {\bar a}^{(\alpha)}_i + \dot A^{(\alpha)}_i )v_i(2t) dv +
 \frac12 \int_B \sum_{\alpha=1}^2 \rho^{(\alpha)} \dot {\bar a}^{(\alpha)}_i
\dot {\bar a}^{(\alpha)}_idv.
\end{array}
 \label{A63}
\end{equation}

Since 
${\bf V} = \{ {\bf v}^{(1)}, {\bf v}^{(2)}, \phi^{(1)},  \phi^{(2)} \} \in \hat {\bf W}_1(B)
\times\hat {\bf W}_1(B)
\times \hat W_1(B)
\times \hat W_1(B)$, from (\ref{energia}), (\ref{A51})  and (\ref{A57}), we deduce that
\begin{equation}
\displaystyle \int_B \sum_{\alpha=1}^2 v^{(\alpha)}_i(s)v^{(\alpha)}_i(s) dv \le \frac{2}{m_1} \int_B \sum_{\alpha=1}^2 W({\bf V}
(s)) dv \le \frac{2}{m_1} {\cal E}(0). \label{A64}
\end{equation}

Taking into account of the equations (\ref{A63}) and (\ref{A64}), we have
\begin{equation}
\displaystyle \lim_{t \to \infty}  \frac{1}{4t} \int_B \sum_{\alpha=1}^2 \rho^{(\alpha)} \dot u^{(\alpha)}_i(0)
 u^{(\alpha)}_i(2t) dv = \frac12 \int_B \sum_{\alpha=1}^2 \rho^{(\alpha)}
\dot {\bar a}^{(\alpha)}_i \dot {\bar a}^{(\alpha)}_i dv .
 \label{A65}
\end{equation}
With the aid of the equations (\ref{A61}) and  (\ref{A65}), we conclude that
\begin{equation}
\begin{array}{l}
\displaystyle
 \lim_{t \to \infty} {\cal K}_{C}(t) = \lim_{t \to \infty}
S_{C}(t) + \frac12 \int_B \sum_{\alpha=1}^2 \rho^{(\alpha)} 
\dot a_i^{(\alpha)} \dot a_i^{(\alpha)} dv . 
\end{array}
 \label{A651}
\end{equation}
Moreover,  the equations (\ref{A58}) and (\ref{A651})  lead to (\ref{A54}).

\bibliographystyle{plain}

\end{document}